\documentclass[prb,preprint,showpacs]{revtex4}
\usepackage{epsf}

\begin{document}

\title {Modulated structure in the martensite phase of Ni$_{1.8}$Pt$_{0.2}$MnGa: a neutron diffraction study}  

\author{Sanjay Singh,$^{1}$  K. R. A. Ziebeck,$^2$ E. Suard,$^3$ P. Rajput,$^4$ S. Bhardwaj,$^1$  A. M. Awasthi,$^1$ and S. R. Barman$^{1*}$}
\affiliation{$^1$UGC-DAE Consortium for Scientific Research, Khandwa Road, Indore, 452001,  India.}
\affiliation{$^2$Department of Physics, Cavendish Laboratory, University of Cambridge, CB3 0HE, UK.}
\affiliation{$^3$Institut Laue-Langevin, BP 156, 38042 Grenoble Cedex 9, France.}
\affiliation{$^4$European Synchrotron Radiation Facility, 6 rue Jules Horowitz, F-38000 Grenoble, France.}

\begin{abstract}

$7M$ orthorhombic modulated structure in the martensite phase of Ni$_{1.8}$Pt$_{0.2}$MnGa is reported by powder neutron diffraction study, which indicates that it is likely to exhibit magnetic field induced strain. The change in the unit cell volume is less than 0.5\% between the austenite and martensite phases, as expected for a volume conserving  martensite transformation.
~The magnetic structure analysis shows that the magnetic moment in the martensite phase is higher  compared to Ni$_2$MnGa, which is in good agreement with magnetization measurement. 

\end{abstract}


\maketitle

Ni$_2$MnGa is a ferromagnetic Heusler alloy, which shows a large magnetic field induced strain (MFIS) and fast actuation in the martensite phase.\cite{Ullakko96,Murray00,Sozinov02} These properties make  Ni$_2$MnGa a material with high potential for application as magnetic actuators. However, brittleness and low transition temperature of this material has necessitated the search for new alloys  with similar MFIS, but with higher transition temperatures and ductility.\cite{Pons08} 
~In the Ni-Mn-Ga family, an increased magnetic transition temperature ($T_C$) of 588\,K as well as 4\% MFIS has been reported for Mn$_2$NiGa.\cite{Liu05} Generally, MFIS is observed in structures that exhibit modulation, since that leads to lower twinning stress.\cite{Murray00} The modulation can be described as a shuffling of the (110) planes along the [1$\overline{1}$0] direction.\cite{Webster84,Brown02,Righi06} A modulated structure in Mn$_2$NiGa has been recently reported by x-ray diffraction study.\cite{SSingh10}  In the case of Ga$_2$MnNi, although $T_C$ is lower than Ni$_2$MnGa, a large martensitic start temperature (M$_s$) and modulated structure have been observed.\cite{Barman08,SSingh11}  
~Other ferromagnetic shape memory alloys such as  Ni-Mn-Al,\cite{Fujita00} Ni-Mn-Fe-Ga,\cite{Wang06, Koho04} and Ni-Fe-Ga-Co\cite{Morito05} showing MFIS with improved ductility have been reported. 
~A modulated structure  has also been reported for off-stochiometric Ni-Mn-In compositions.\cite{Krenke07,Brown11}
~However, although the above mentioned materials exhibit MFIS, the magnitude is much smaller compared to Ni$_2$MnGa (10\%). 
~Of late, {\it ab-initio} density functional theory (DFT) has played an important role in predicting new ferromagnetic shape memory alloys.\cite{Barman08,Chakrabarti09}   
~Taking cue from an earlier experimental work,\cite{Wuttig04} a recent DFT study  has put forward Pt doped Ni$_2$MnGa to be an alternative to Ni$_2$MnGa.\cite{Entel11} The theoretical estimate of maximum MFIS is about 14\% that is higher than Ni$_2$MnGa.\cite{Entel11} 
~In this letter, we report  the existence of a modulated structure in the martensite phase of Ni$_{1.84}$Pt$_{0.2}$MnGa$_{0.96}$ from neutron diffraction studies, which is strongly suggests that it would exhibit MFIS. Our analysis shows that the magnetic moment in the  martensite phase is higher than Ni$_2$MnGa.

The specimen has been prepared by melting appropriate quantities of Ni, Pt, Mn and Ga of 99.99\% purity in an arc furnace. Less than  1\% weight loss was observed after melting. The ingot was then annealed at 1173\,K for 3 days for homogeneization and then slow cooled to room temperature. X-ray diffraction at room temperature showed a single phase $L2_1$ structure, as expected for the austenite phase. The composition determined by energy dispersive analysis of x-rays (EDAX) was done using scanning electron microscope with Oxford detector model turned out to be Ni$_{1.84}$Pt$_{0.2}$MnGa$_{0.96}$,  which we nominally refer to as Ni$_{1.8}$Pt$_{0.2}$MnGa henceforth in the manuscript. ~Powder neutron diffraction 
~was  performed using  neutrons of wavelength 1.59\AA~  at the D2B neutron diffractometer in ILL, Grenoble. The specimen was placed in a cylindrical vanadium cylinder inside a furnace for recording diffraction patterns at 450\,K and 300\,K and inside a cryofurnace for the 230\,K measurement.  Rietveld analysis of the neutron data were carried out using the  FULLPROF software package.\cite{FullProf00} 

~  To obtain the  transition temperatures differential scanning calorimetry (DSC) was performed using TA Instruments MDSC model 2910 at a scan rate of 10$^{\circ}$/minute.  Magnetization measurements were performed using a MPMS XL5 superconducting quantum interference device (SQUID) magnetometer. 
~From DSC, we obtain the martensite start ($M_s$), martensite finish ($M_f$), austenite start ($A_s$) and austenite finish ($A_f$) temperatures to be 285\,K, 260\,K, 274\,K and 305\,K, respectively  (Fig. 1(a)). These  temperatures are slightly different from the values reported earlier,\cite{Entel11,Entel12,note} which is possibly related to difference in composition. It is well known that in Ni-Mn-Ga the martensite transition temperatures vary sensitively with composition.\cite{Liu05,Banik09} An important observation is that M$_s$ (285\,K) of this material is close to room temperature, in contrast to Ni$_2$MnGa (200\,K), which makes it more attractive for practical applications. The latent heat of transformation turns out to be 1.09\,kJoule/mole, which is similar to Ni$_2$MnGa.\cite{Banik07} The signature of the magnetic transition at $T_C$=\,352$\pm$2\,K is observed in both heating and cooling curves of DSC. The $M(H)$ hysteresis loops  for the martensite (250\,K) and austenite (300\,K) phases  show that both the phases are ferromagnetic with magnetic moments of 3.55$\mu_B$/f.u. and 2.8$\mu_B$/f.u., respectively (Fig. 1(b)). 


\begin{figure}[htb]
\epsfxsize=65mm
\epsffile{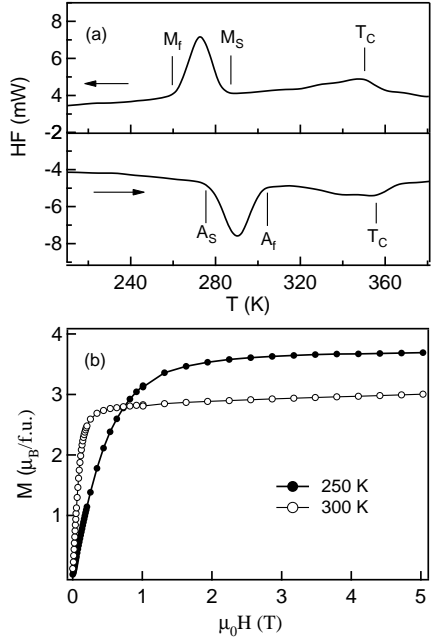}
\caption{(color online) (a) DSC heating and cooling curves and (b)  $M(H)$ hysteresis loops at 300\,K  and 250\,K for  Ni$_{1.8}$Pt$_{0.2}$MnGa.}  
\end{figure}

Fig.~2 shows that all the neutron diffraction peaks in the austenite phase obtained at 450\,K (paramagnetic) and 300\,K (ferromagnetic, $T_C$=\,355\,K) can be indexed well using the cubic L2$_1$ structure. The lattice parameter $a$ is 5.88 at 450\,K, while it is 5.852\AA~ at 300\,K. Compared to Ni$_2$MnGa ($a$=\,5.825 at 300\,K)\cite{Brown02}, the lattice constant is larger in Ni$_{1.8}$Pt$_{0.2}$MnGa. Increase in the volume of the unit cell with Pt doping in Ni$_2$MnGa has been predicted by DFT calculations\cite{Entel11}.
~The Rietveld refinement was performed using space group Fm$\overline{3}$m, where  Ni and Pt atoms occupy the 8$c$ (0.25,\,0.25,\,0.25) position, while Mn and Ga are at 4$a$ (0,\,0,\,0) and 4$b$ (0.5,\,0.5,\,0.5) Wyckoff positions, respectively. The significantly different coherent nuclear scattering amplitudes of Ni (10.3 fm), Mn (-3.73 fm), Ga (7.29 fm) and Pt (9.6 fm) allows the determination of the occupancies of each site. 
~In order  to determine the atomic structure in the ferromagnetic state (300\,K), the  occupancies were also refined  by fitting the diffraction pattern above 2$\theta$=\,60$^\circ$, where the magnetic contribution in the peak intensity is negligible. The refined occupancies at 300\,K were similar to 450\,K,  which rules out the possibility of temperature induced disorder effect at 450\,K. After determining the site occupancies (Table~I), the refinement of the diffraction pattern in 2$\theta$ range 20$^\circ$-140$^\circ$~was carried out including a ferromagnetic moment on the manganese atoms characterized by a Mn$^{2+}$ form factor. The refinement results clearly show that the Pt atoms occupy only the Ni site and Ni-Mn or Mn-Ga anti-site disorder are absent. Antisite disorder was also reported to be absent in   Ni$_2$MnGa.\cite{Brown02} 
~
~The magnetic moment of Ni$_{1.8}$Pt$_{0.2}$MnGa at 300\,K turns out to be 2.44(06)~$\mu_B$ from our analysis and this is close to the  the value (2.8~$\mu_B$) obtained from magnetization measurement (Fig.~1b). 

 \begin{figure}[htb]
 \epsfxsize=70mm
 \epsffile{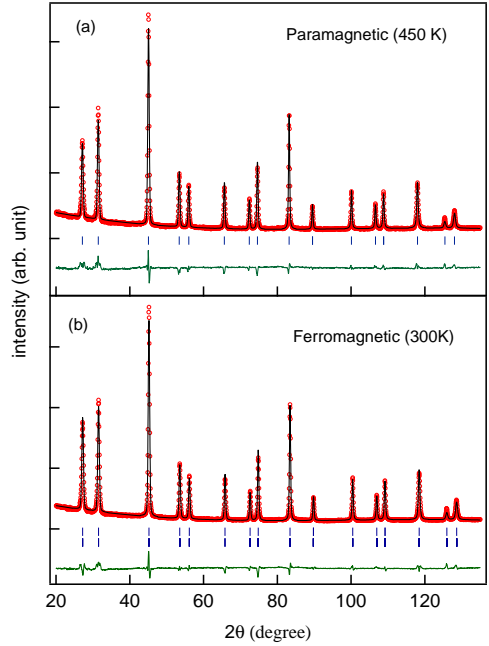}
\caption{(color online)  Rietveld fitting of the powder neutron diffraction pattern in the austenite phase in the (a)  paramagnetic state at 450\,K and (b)  ferromagnetic state at 300\,K for Ni$_{1.8}$Pt$_{0.2}$MnGa.  The experimental data, fitted curves and the residue are shown by red open circles, black line and green line, respectively. The upper and lower rows of blue ticks represent the crystal and magnetic Bragg peak positions, respectively.}
\end{figure}           

\begin{figure}[htb]
\epsfxsize=90mm
\epsffile{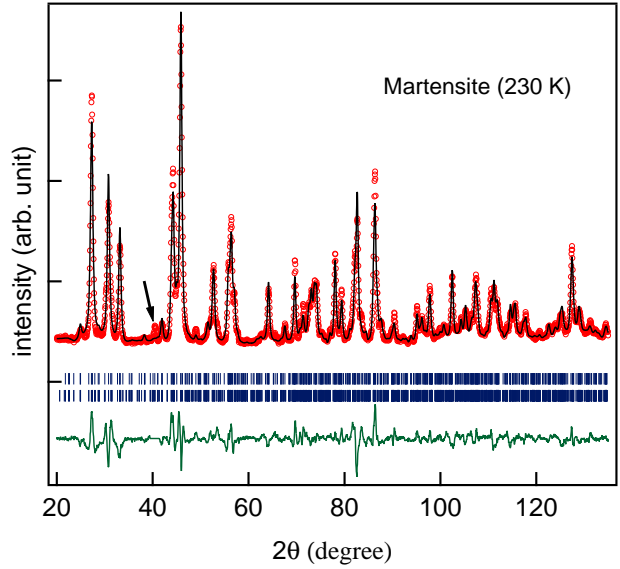}
\caption{(color online)  Rietveld fitting of the powder neutron diffraction pattern of Ni$_{1.8}$Pt$_{0.2}$MnGa in the martensite phase at 230\,K. The symbols have the same meaning as in Fig.~2.
~The arrow indicates a peak due to Aluminum (see text).}
\end{figure}

\begin{table}[htb] 
 \begin{center} 
 \caption{\footnotesize Parameters obtained from the refinement of the neutron diffraction pattern of Ni$_{1.8}$Pt$_{0.2}$MnGa in the austenite phase in paramagnetic (450\,K) and ferromagnetic (300\,K) state.}
 \begin{tabular}{l  c   c  c c} \\\hline\hline
Temperature     &  &  \hspace{0.70 cm}&\hspace{0.15 cm} 450\,K   \hspace{0.30 cm} &\hspace{0.50 cm}300K\\
Space group     &  & \hspace{0.70 cm} &Fm$\overline{3}$m   \hspace{0.15 cm}&\hspace{0.50 cm}Fm$\overline{3}$m        \\
Lattice parameter &  &\hspace{0.70 cm} &\hspace{0.15 cm}5.884\AA  \hspace{0.15 cm}& \hspace{0.50 cm}5.852\AA  \\
Cell volume (\AA$^3$)& &\hspace{0.70 cm}& \hspace{0.15 cm}203.70 \hspace{0.15 cm}& \hspace{0.20 cm}200.43 \\ 
$\chi^2$       && \hspace{0.70 cm}& \hspace{0.15 cm}1.5 \hspace{0.15 cm} & \hspace{0.50 cm}2      \\\hline
Atom       & site& \hspace{0.70 cm}&  \hspace{0.15 cm} Occ.  \hspace{0.15 cm} &  \hspace{0.50 cm}Moment ($\mu_B$)   \\\hline

Ni  &8$c$ & \hspace{0.70 cm}& \hspace{0.15 cm}   1.8          \hspace{0.15 cm} &    \hspace{0.50 cm}       -    \\
Pt &8$c$ & \hspace{0.70 cm}& \hspace{0.15 cm}     0.2        \hspace{0.15 cm} &    \hspace{0.50 cm}       -    \\
Mn          &4$a$ &\hspace{0.70 cm}&   \hspace{0.15 cm}   1         \hspace{0.15 cm} &      \hspace{0.50 cm}     2.44(06)   \\  							
Ga         &4$b$& \hspace{0.50 cm}&   \hspace{0.05cm}     1     \hspace{0.15 cm}&     \hspace{0.50 cm}    -         \\\hline\hline   

\end{tabular} 
\end{center}
\end{table} 

\begin{figure*}[htb]
\epsfxsize=150mm
\epsffile{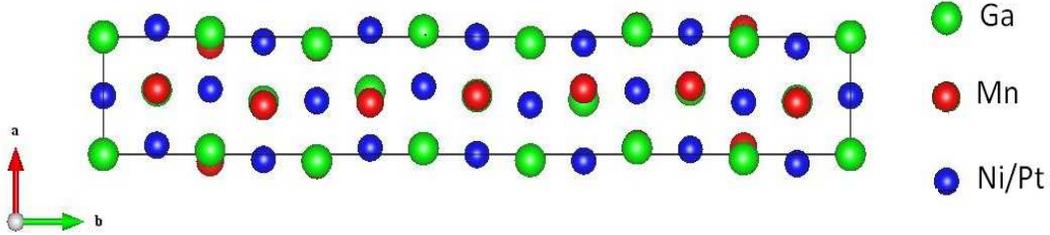}
\caption{(color online)  The orthorhombic unit cell of Ni$_{1.8}$Pt$_{0.2}$MnGa (from Table~II) projected in the $a-b$ plane highlights the modulation effect.}  
\end{figure*}

\begin{table}[htb] 
\begin{center} 

\caption{\footnotesize Lattice parameters, atomic positions ($x, y, z$) and  magnetic moments ($\mu_B$) of Ni$_{1.8}$Pt$_{0.2}$MnGa in the martensite phase at 230\,K. 
}
\begin{tabular}{l       c       c       c      c      r} \\\hline\hline
Crystal system \hspace{0.30cm}  & \hspace{0.50cm}& \hspace{0.25cm} Orthorhombic\\
Space group    \hspace{0.30cm} & \hspace{0.50cm}&  \hspace{0.25cm}Pnnm\\
Cell (\AA)      &  $a$=\,4.261, & $b$=\,29.604,   &   \hspace{0.50cm} $c$=\,5.583 
\end{tabular}  
\begin{tabular}{l       c       c       c       c      c } \hline
Atom        &        Wych.       &   $x$               &      \hspace{0.30  cm}    $y$   \hspace{0.30  cm}      &    \hspace{0.30  cm} $z$   \hspace{0.10  cm} &  Mom. ($\mu_B$)    \\ \hline
Mn1      &         2a          &                 0                  &          0            &      0  &      3.60(1) \\
Ga1      &         2b          &                 0                 &           0            &      0.5  &         \\ 
0.9Ni1+0.1Pt1      &         4f          &       0.5                  &        0.0          &      0.25  &          \\
Mn2      &         2a          &                 0.055(13)           &         1/7            &    0 &      3.60(1)\\
Ga2      &         2b          &                 -0.038(9)           &         1/7             &   0.5 &       \\
0.9Ni2+0.1Pt2    & 4f          &                 0.443(2)             &        1/7             &   0.25  &        \\ 
Mn3      &         4g          &                 0.087(13)                 &   2/7              &  0  &     3.60(1) \\
Ga3      &         4g          &                 0.058(7)               &      2/7           &     0.5  &          \\ 
0.9Ni3+0.1Pt3     &8h          &                 0.540(2)               &      2/7            &    0.25  &       \\
Mn4      &         4g          &                 -0.057(11)            &       3/7             &   0 &       3.60(1) \\         
Ga4      &         4g          &                 -0.050(7)            &        3/7          &      0.5 &        \\
0.9Ni4+0.1Pt4     &8h          &                 0.420(2)                 &    3/7             &   0.25   &        \\\hline

    &           &                $\chi^2$ = 2.70                 &                  &       &        \\\hline\hline
    
\end{tabular} 
\end{center}
\end{table} 

Ni$_{1.8}$Pt$_{0.2}$MnGa is expected to transform to the martensite phase below the martensite finish temperature of M$_f$=\,260\,K.  
~The diffraction pattern at 230\,K (Fig.~3) that depicts the martensite phase is found to be completely different from the austenite phase (Fig.~2) with the occurrence of many extra peaks.  
~The absence of  peaks related to the L2$_1$ phase confirms that the transformation to the martensite phase is complete. In order to analyze this diffraction pattern, initially Lebail fitting trials using the tetragonal, orthorhombic and monoclinic unit cells were performed. But these failed to account for all the Bragg peaks. This suggests the possibility of the existence of a modulated phase. Since the Pt doping is small (10\%) in Ni$_{1.8}$Pt$_{0.2}$MnGa, an attempt was made to fit the pattern using the orthorhombic unit cell and space group $Pnnm$ that has been proposed for Ni$_2$MnGa.\cite{Brown02}  The starting unit cell parameters were taken as $a_{ortho}$= (${1\over\sqrt{2}}$)$a_{cubic}$, $b_{ortho}$=  (${7\over\sqrt{2}}$)$a_{cubic}$ and $c_{ortho}$= $a_{cubic}$. All the Bragg peaks could be indexed  by using this orthorhombic unit cell. The only peak that could not be accounted for at 2$\theta$=\,40.5$^{\circ}$  is due to aluminum in the cryofurnace wall.
~So, the 2$\theta$ range of this peak was excluded during the refinement. The lattice parameters obtained from the refinement are  $a$=\,4.261~\AA, $b$=\,29.604~\AA~ and $c$=\,5.583~\AA, which are larger than Ni$_2$MnGa.\cite{Brown02,Righi06} However, the relation between $b$ and $a$ is found to be $b$$\approx$\,7$a$, which indicates a seven-fold increase in the unit cell along $b$. A similar relation between $b$ and $a$ has been reported for  Ni$_2$MnGa and ascribed to the $7M$ modulated structure.\cite{Brown02,Righi06,Ranjan06} The atomic structure refinement was performed using the Rietveld method and the refined atomic positions are shown in Table~II.  The modulation waves for all the atoms (Ni/Pt, Mn and Ga) are clearly observed in Fig.~4. If compared to Ni$_2$MnGa,\cite{Brown02, Righi06} the amplitude of modulation for Mn and Ga is larger,  while it is smaller for Ni.  The unit-cell volume of the martensite phase (704.25~\AA$^3$) is within 0.5\% of that of a comparable austenite cell volume given by 7$\times$$a_{aus}^3$/2 (701.42~\AA$^3$). Thus the unit-cell volume of the two phases is very similar, which is a necessary condition for a shape memory behaviour. 
~After the Rietveld refinement of the atomic positions and the magnetic structure (Fig.~3), we find that  
~the magnteic moment is mainly confined to the Mn atoms, which carry a ferromagnetic moment directed perpendicular to the long axis. The magnetic moment  is found to be 3.6~$\mu_B$. This is in excellent agreement with  the value (3.55~$\mu_B$) obtained from the magnetization measurement,  as shown in the Fig.~1b. 



We thank P. Entel, A. Chakrabarti 
~ P. J. Brown and R. Ranjan for fruitful suggestions and discussions. E. V. Sampathkumaran and K. Mukherjee are thanked for useful discussions and for providing the magnetization data. B. A. Chalke is thanked for providing the EDAX data. Funding from the Max Planck Partner Group project; Department of Science and Technology, Government of India;  and Institut Laue-Langevin, France is gratefully acknowledged. S. Singh is thankful to the Council for Scientific and Industrial Research for research fellowship.\\  

\noindent $^{*}$barmansr@gmail.com


\begin{thebibliography}{}
\bibitem{note} Experimental $A_F$ value (330\,K) noted in Ref.\onlinecite{Entel11} is a misprint, the actual value is 300\,K, as given in Ref.\onlinecite{Entel12}.
\end{thebibliography}
\end{document}